\documentclass{article}
\usepackage{spconf,amsmath,amssymb,graphicx,booktabs}
\usepackage[hidelinks]{hyperref}
\title{MPASCNet: Magnitude-Phase-Aware Adaptive Spectral Compression for Compressed Music Restoration}

\name{Xinlong Deng$^1$, Yu Xia$^1$, Qiuqiang Kong$^2$, Yin Cao$^{3}$, Jie Jiang$^1$}
\address{
$^1$College of Artificial Intelligence, China University of Petroleum, Beijing\\
$^2$Department of Electronic Engineering, The Chinese University of Hong Kong, Hong Kong, China\\
$^{3}$ Institute of Acoustics, Chinese Academy of Sciences\\
2026216101@student.cup.edu.cn, qqkong@ee.cuhk.edu.hk, yin.k.cao@gmail.com, jiangjie@cup.edu.cn
}

\begin{document}
% \ninept
\maketitle

\begin{abstract}
Compressed music restoration aims to recover high-fidelity audio from lossy codecs such as MP3 by restoring missing spectral content and reducing coding artifacts. Recently, generative adversarial networks and band-split models have made progress in compressed music restoration. However, existing methods either operate without explicitly modeling phase or rely on predefined frequency partitions, which may limit restoration performance and the use of input-dependent frequency information. To address these issues, we propose MPASCNet, a Magnitude-Phase-Aware Adaptive Spectral Compression network. Instead of using fixed bands, MPASCNet combines magnitude and phase features to guide cross-attention, adaptively compressing full-resolution spectral features into compact tokens. These tokens are then processed by a spectro-temporal backbone to model spectral-temporal dependencies. Finally, separate decoder branches reconstruct the full-resolution magnitude and wrapped phase spectra. Experiments on MP3 restoration show that MPASCNet outperforms the Apollo and AEROMamba baselines across multiple restoration metrics while using only 6.49 M parameters. The code is available at https://github.com/xinghour/MPASCNet.
\end{abstract}


% ============================================================================
\vspace{-2mm}
\section{Introduction}
\vspace{-2mm}
\label{sec:intro}

Compressed music restoration aims to recover high-fidelity music from lossy-compressed signals such as MP3. MP3 reduces bitrate through perceptual coding and quantization, which at low bitrates can cause bandwidth reduction, spectral clipping, pre-echo, and other audible artifacts~\cite{herre2014perceptual,liu2008compression,brandenburg2013perceptual}. Recovering lost or distorted content is challenging because music contains complex spectro-temporal structures involving harmonics, transients, multiple instruments, and vocals~\cite{li2025apollo}. The task is also practically relevant for improving low-bitrate music and restoring legacy recordings when higher-quality sources are unavailable~\cite{lattner2021stochastic,moliner2023behmgan}.

Learning-based approaches have been widely explored for compressed music restoration.
TF-LSTM~\cite{deng2020tflstm} models temporal and frequency dynamics to reconstruct high-quality magnitude spectra from compressed inputs, while reusing the phase of the compressed signal for waveform reconstruction.
SR-GAN~\cite{lattner2021stochastic} introduces adversarial learning for the restoration of heavily compressed music, learning to restore degraded complex spectrograms toward their high-quality counterparts.
More recently, Apollo~\cite{li2025apollo} introduces band-sequence modeling with predefined subbands and interleaved spectral-temporal modeling.
AEROMamba$_{\mathrm{P}\bar{\mathrm{S}}}$~\cite{Abreu2026efficient}
explores Mamba-based enhancement of heavily compressed MP3 audio with a
differentiable psychoacoustic loss. These studies demonstrate the effectiveness of spectral reconstruction, spectro-temporal modeling, and perceptually motivated learning for compressed music restoration.

However, explicit magnitude-phase modeling remains underexplored in
compressed music restoration. In speech enhancement, MP-SENet~\cite{lu2023mpsenet} has demonstrated the effectiveness of explicitly estimating magnitude and wrapped phase through dedicated branches. Meanwhile, Apollo~\cite{li2025apollo} follows the band-split paradigm introduced by BSRNN~\cite{luo2023bsrnn}, organizing the spectrum with predefined frequency partitions. In music source separation, Spectral Feature Compression (SFC)~\cite{saijo2026sfc} replaces band-split with cross-attention-based input-adaptive aggregation, compressing full-resolution features over frequency bins into compact spectral tokens.

Inspired by these advances in source separation and speech enhancement, we propose MPASCNet, a Magnitude-Phase-Aware Adaptive Spectral Compression network that extends input-adaptive spectral compression to the restoration setting by incorporating explicit magnitude-phase modeling into both spectral compression and reconstruction. Magnitude and wrapped-phase features jointly guide cross-attention to form compact spectral tokens, which are processed by a spectro-temporal backbone and then expanded through magnitude- and phase-specific queries for full-resolution spectral reconstruction. Experiments on a non-overlapping MedleyDB evaluation set show that MPASCNet consistently outperforms both Apollo and AEROMamba$_{\mathrm{P}\bar{\mathrm{S}}}$ across multiple restoration metrics, while using only 6.49 M parameters.

% ============================================================================
\begin{figure*}[t]
  \centering
  \includegraphics[width=1\textwidth]{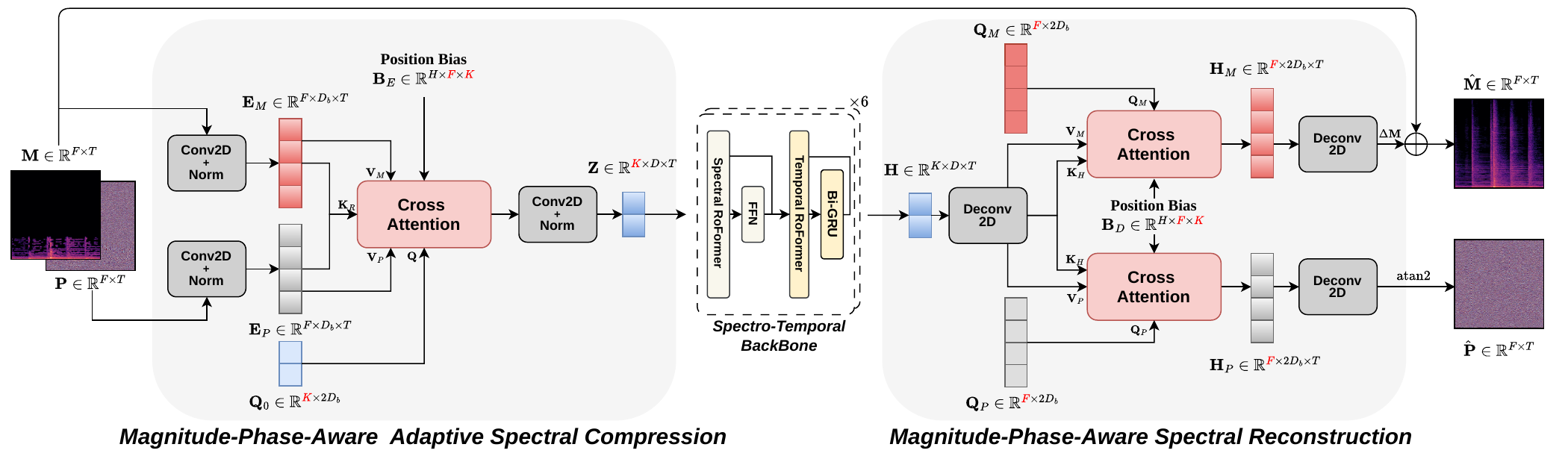}
  \caption{The overview structure of MPASCNet.}
  \label{fig:overview}
\end{figure*}

\section{MPASCNet}
\label{sec:method}

\vspace{-3mm}
\subsection{Overall Pipeline}
\vspace{-1mm}

Fig.~\ref{fig:overview} illustrates the architecture of MPASCNet, comprising magnitude-phase-aware adaptive spectral compression, spectro-temporal backbone, and magnitude-phase-aware spectral reconstruction. Stereo channels share the same network. We describe the processing of a single channel below. Given a compressed signal $x \in \mathbb{R}^{L}$, STFT produces the complex spectrum $X \in \mathbb{C}^{F\times T}$, represented by its power-compressed magnitude $M$ and wrapped phase $P$, where power compression limits the magnitude dynamic range. The adaptive spectral compression module jointly exploits magnitude and phase features to guide cross-attention, adaptively aggregating the \(F\) frequency bins into \(K\) compact spectral tokens \(Z \in \mathbb{R}^{K\times D\times T}\) without predefined frequency partitions. The backbone models spectral and temporal dependencies over these tokens to produce \(G \in \mathbb{R}^{K\times D\times T}\). The reconstruction module then expands \(G\) back to full frequency resolution through separate magnitude and phase branches, producing \(\hat{M}\) and \(\hat{P}\), which are combined into \(\hat{X}\) and transformed via inverse STFT to obtain \(\hat{x}\).

\vspace{-3mm}
\subsection{Magnitude-Phase-Aware Adaptive Spectral Compression}
\vspace{-1mm}

Following the cross-attention formulation of SFC~\cite{saijo2026sfc}, $K$ learnable spectral queries aggregate the $F$ frequency bins into $K$ compact tokens by attending over the full frequency axis. Unlike SFC, which operates on generic real-imaginary spectral features, MPASCNet jointly uses magnitude and phase representations to determine the frequency-to-token aggregation, allowing the adaptive aggregation to incorporate both spectral-energy and complementary phase cues.

For each time frame, the cross-attention operates along the frequency axis. Given $M,P\in\mathbb{R}^{F\times T}$, two convolutional embeddings with RMSNorm produce $E_M,E_P\in\mathbb{R}^{F\times D_b\times T}$, where $D_b$ denotes the branch embedding dimension. They are concatenated along the feature dimension to form $R\in\mathbb{R}^{F\times 2D_b\times T}$. Together with $K$ learnable spectral queries $Q_0\in\mathbb{R}^{K\times 2D_b}$, linear projections generate the query $Q$ from $Q_0$ and the key $K_R$ from $R$. Meanwhile, $E_M$ and $E_P$ are independently projected into $V_M$ and $V_P$, which are concatenated as the value $V$. Thus, the frequency-to-token aggregation is jointly conditioned on magnitude and phase, while independent value projections allow the two representations to retain their feature transformations before aggregation.

To provide a frequency prior for the queries, we use a learnable positional bias $B_E\in\mathbb{R}^{H\times K\times F}$, where $H$ denotes the number of attention heads. Compared with SFC’s low-frequency-emphasized initialization, we use approximately equal-width bands for $B_E$, assigning residual bins to the last band to reserve more capacity for mid- and high-frequencies susceptible to compression-induced loss.

The cross-attention is computed as
\begin{equation}
\operatorname{CrossAttn}(Q,K_R,V)
=
\operatorname{Softmax}
\left(
\frac{QK_R^{\top}}{\sqrt{d}} + B_E
\right)V,
\end{equation}
where $d$ is the per-head feature dimension and the softmax is applied over the $F$ frequency bins. Each query adaptively aggregates full-resolution value features into one compact spectral token. The outputs of all attention heads are concatenated along the feature dimension, followed by a convolutional projection with RMSNorm to obtain $Z\in\mathbb{R}^{K\times D\times T}$.

\vspace{-3mm}
\subsection{Spectro-Temporal Backbone}
\vspace{-1mm}

Spectro-temporal modeling has been widely adopted in time-frequency audio models, including TF-GridNet~\cite{wang2023tfgridnet} and BS-RoFormer~\cite{lu2024bsroformer}, to capture dependencies across the two axes. Following this paradigm, we model the compressed representation $Z\in\mathbb{R}^{K\times D\times T}$, where adaptive spectral tokens replace predefined subbands.

Each block first applies RoFormer attention along the spectral-token axis followed by an FFN, and then temporal RoFormer attention followed by a bidirectional GRU~\cite{cho2014gru} to complement global attention with recurrent modeling of temporal continuity. Stacking these blocks produces $G\in\mathbb{R}^{K\times D\times T}$ for spectral reconstruction.

\subsection{Magnitude-Phase-Aware Spectral Reconstruction}
\vspace{-2mm}

For reconstruction, the compact representation is expanded from $K$ spectral tokens back to $F$ frequency bins. To preserve the explicit magnitude-phase decomposition, we use separate magnitude and phase branches rather than a shared complex-spectrum mapping.

Given the backbone output $G\in\mathbb{R}^{K\times D\times T}$, a convolutional projection with RMSNorm maps it to $\tilde{G}\in\mathbb{R}^{K\times 2D_b\times T}$. Two sets of learnable full-resolution queries, $Q_M$ and $Q_P$, are used for magnitude and phase reconstruction, respectively. Each query corresponds to one output frequency bin and attends over the $K$ spectral tokens. Both branches use a shared key $K_G$ derived from $\tilde{G}$, while $V_M$ and $V_P$ are projected independently for magnitude and phase reconstruction. This allows both branches to decode from the same compressed spectral context, while branch-specific queries and values retain the different reconstruction characteristics of magnitude and phase.
A learnable positional bias $B_D\in\mathbb{R}^{H\times F\times K}$, initialized from the transpose of $B_E$, guides the expansion from spectral tokens back to frequency bins. For branch $r\in\{M,P\}$, the branch-specific cross-attention is computed as
\begin{equation}
\operatorname{CrossAttn}(Q_r,K_G,V_r)
=
\operatorname{Softmax}
\left(
\frac{Q_rK_G^\top}{\sqrt{d}} + B_D
\right)V_r .
\end{equation}

The two branches consequently produce separate magnitude and phase features at full frequency resolution. The magnitude branch predicts a residual correction $\Delta M$, yielding $\hat{M}=M+\Delta M$. The phase branch predicts a two-dimensional phase representation rather than directly regressing the wrapped angle, from which $\hat{P}$ is recovered using $\operatorname{atan2}$. The resulting $\hat{M},\hat{P}\in\mathbb{R}^{F\times T}$ are combined to construct the restored complex spectrum $\hat{X}$.

\vspace{-4mm}
\subsection{Training Objective}
\vspace{-1mm}

We use a multi-domain objective that jointly constrains magnitude, phase, complex spectrum, STFT consistency, and waveform reconstruction. $\mathcal{L}_{\mathrm{mag}}$, $\mathcal{L}_{\mathrm{com}}$, and $\mathcal{L}_{\mathrm{stft}}$ are computed using MSE for compressed magnitude, compressed complex spectrum, and STFT consistency, respectively, while $\mathcal{L}_{\mathrm{time}}$ uses the $\ell_1$ waveform distance.

For phase reconstruction, we adopt the anti-wrapping phase objective of MP-SENet~\cite{lu2023mpsenet}, which jointly constrains instantaneous phase (IP), group delay (GD), and instantaneous angular frequency (IAF). Let $f_{\mathrm{AW}}(\cdot)$ denote the anti-wrapping function defined in~\cite{lu2023mpsenet}. The phase objective is
\begin{equation}
\mathcal{L}_{\mathrm{pha}}
=
\sum_{d\in\{0,f,t\}}
\mathbb{E}
\left[
f_{\mathrm{AW}}
\left(
\Delta_d(P^*-\hat{P})
\right)
\right],
\end{equation}
where $\Delta_0$ denotes the identity operation, while $\Delta_f$ and $\Delta_t$ denote first-order differences along the frequency and temporal axes, corresponding to IP, GD, and IAF, respectively.

The overall objective is
\begin{equation}
\begin{aligned}
\mathcal{L}
={}&
\lambda_{\mathrm{mag}}\mathcal{L}_{\mathrm{mag}}
+
\lambda_{\mathrm{pha}}\mathcal{L}_{\mathrm{pha}}
+
\lambda_{\mathrm{com}}\mathcal{L}_{\mathrm{com}}
\\
&+
\lambda_{\mathrm{stft}}\mathcal{L}_{\mathrm{stft}}
+
\lambda_{\mathrm{time}}\mathcal{L}_{\mathrm{time}}.
\end{aligned}
\end{equation}
where $\lambda_{\mathrm{mag}}$, $\lambda_{\mathrm{pha}}$, $\lambda_{\mathrm{com}}$, $\lambda_{\mathrm{stft}}$, and $\lambda_{\mathrm{time}}$ are the loss weights, set to $1.2$, $0.3$, $0.1$, $0.1$, and $0.2$, respectively.

% ============================================================================
\vspace{-3mm}
\section{Experimental Setup}
\vspace{-1mm}
\label{sec:setup}

\vspace{-2mm}
\subsection{Datasets}
\vspace{-1mm}

We use MUSDB18-HQ~\cite{rafii2017musdb18} and MoisesDB~\cite{pereira2023moisesdb} for training, following the data preparation protocol of Apollo~\cite{li2025apollo}. After source activity detection, 3-s excerpts are sampled from active stems, randomly mixed, and degraded by MP3 compression with the bitrate randomly selected from 24, 32, 48, 64, 96, and 128 kbps.
Since Apollo does not specify the exact composition of its validation and test sets, we construct a non-overlapping evaluation set from MedleyDB~\cite{bittner2014medleydb}. We remove the 46 tracks shared with MUSDB18-HQ, leaving 150 tracks for evaluation. We consider Multi-Stem using the original mixtures, Single-Stem using individual processed stems, and Vocal using vocal stems. For each condition, 500 3-s excerpts are deterministically sampled across tracks and degraded using the same MP3 bitrate scheme.

\vspace{-3mm}
\subsection{Training Configuration}
\vspace{-1mm}
\label{ssec:training}

Following Apollo~\cite{li2025apollo}, all audio is processed at 44.1 kHz using an STFT with an 882-sample Hann window and a hop size of 441 samples. We set the backbone feature dimension to $D=128$ and stack six spectro-temporal blocks with eight attention heads. The adaptive spectral compression module uses an inner dimension of 64, four attention heads $d=16$, and $K=80$ compressed spectral tokens. The model is trained with AdamW using a batch size of 2, an initial learning rate of $10^{-4}$, $(\beta_1,\beta_2)=(0.8,0.99)$, and an exponential learning-rate decay of 0.99 per epoch. Early stopping is applied when the validation SI-SNR does not improve for 20 consecutive epochs. All experiments are conducted on a single NVIDIA GeForce RTX 4090 GPU.

\vspace{-3mm}
\subsection{Baselines and Evaluation Metrics}
\vspace{-1mm}
\label{ssec:evaluation metrics}
We compare MPASCNet against two baseline methods, Apollo~\cite{li2025apollo} and AEROMamba$_{\mathrm{P}\bar{\mathrm{S}}}$~\cite{Abreu2026efficient}. For Apollo, we use the publicly released model weights. For AEROMamba$_{\mathrm{P}\bar{\mathrm{S}}}$, we adapt its original MP3 enhancement setting to our multi-bitrate restoration protocol and retrain it using the same training data.
Restoration quality is evaluated using SDR, SI-SNR, and ViSQOL~\cite{chinen2020visqol}. 
For spectral reconstruction, we report magnitude-weighted phase distance (PD)~\cite{choi2019phaseaware} in degrees and log-spectral distance (LSD)~\cite{moliner2023behmgan}. PD measures the wrapped phase difference weighted by the reference magnitude, while LSD measures the root-mean-square difference between reference and reconstructed log-power spectra. Model efficiency is evaluated by trainable parameters and real-time factor (RTF), reported in units of $10^{-3}$.

% ============================================================================

\begin{table*}[t]
\centering
\caption{Restoration performance on MedleyDB. Scores are reported as
SDR$\uparrow$/SI-SNR$\uparrow$/ViSQOL$\uparrow$.}
\label{tab:main_results}
{%
\setlength{\tabcolsep}{7.0pt}
\renewcommand{\arraystretch}{1.08}
\begin{tabular}{lcccccc}
\toprule
Method & Vocal & Single-Stem & Multi-Stem & Overall & Params (M) & RTF \\
\midrule
MP3 Input
& 15.91/15.80/2.95
& 12.41/9.86/3.18
& 13.45/10.99/2.93
& 13.92/12.22/3.02
& -- & -- \\

AEROMamba$_{\mathrm{P}\bar{\mathrm{S}}}$\cite{Abreu2026efficient}
& 17.04/16.68/3.45
& 16.21/15.45/3.47
& 15.52/15.06/3.55
& 16.26/15.73/3.49
& 18.55 & \textbf{4.10} \\

Apollo\cite{li2025apollo}
& 17.09/16.98/3.56
& 16.03/15.29/3.57
& 16.71/15.82/3.67
& 16.61/15.77/3.60
& 16.54 & 37.54 \\

\textbf{MPASCNet}
& \textbf{18.10/17.94/3.60}
& \textbf{17.47/16.74/3.65}
& \textbf{17.37/17.14/3.74}
& \textbf{17.65/17.27/3.67}
& \textbf{6.49}
& 23.06 \\
\bottomrule
\end{tabular}
}
\end{table*}

\section{Results and Discussion}
\label{sec:results}

\begin{table}[t]
\centering
\caption{Spectral reconstruction analysis on MedleyDB.}
\label{tab:spectral_analysis}
{%
\setlength{\tabcolsep}{4.0pt}
\renewcommand{\arraystretch}{1.08}
\begin{tabular}{lccc}
\toprule
Metric & AEROMamba$_{\mathrm{P}\bar{\mathrm{S}}}$ & Apollo & MPASCNet \\
\midrule
PD $\downarrow$ & 16.62 & 14.88 & \textbf{13.80} \\
LSD $\downarrow$ & 1.12 & 1.04 & \textbf{0.92} \\
\bottomrule
\end{tabular}
}
\end{table}

\vspace{-1mm}
\subsection{Restoration Performance on MedleyDB}
\vspace{-1mm}
\label{ssec:main_results}

Table~\ref{tab:main_results} compares MPASCNet with AEROMamba$_{\mathrm{P}\bar{\mathrm{S}}}$ and Apollo on the MedleyDB evaluation set. We also report the compressed MP3 input without restoration as a reference. MPASCNet consistently achieves the best SDR, SI-SNR, and ViSQOL across Vocal, Single-Stem, and Multi-Stem conditions. Overall, MPASCNet obtains an SDR of 17.65 dB and an SI-SNR of 17.27 dB, improving over Apollo by 1.04 dB and 1.50 dB, respectively. MPASCNet also achieves the highest overall ViSQOL score of 3.67. Moreover, it uses only 6.49 M parameters, substantially fewer than Apollo (16.54 M) and AEROMamba\(_{\mathrm{PS}}\) (18.55 M). In terms of inference speed, MPASCNet achieves a lower RTF than Apollo, although AEROMamba\(_{\mathrm{PS}}\) remains faster.
% AEROMamba$_{\mathrm{P}\bar{\mathrm{S}}}$ achieves the lowest RTF, benefiting from the computational efficiency of its Mamba-based sequence modeling. MPASCNet consistently outperforms both baselines in SDR, SI-SNR, and ViSQOL across Single-Stem, Multi-Stem, and Vocal conditions, while using substantially fewer parameters.

Table~\ref{tab:spectral_analysis} further evaluates spectral reconstruction using PD and LSD. MPASCNet achieves the lowest PD and LSD among the compared models, providing complementary evidence of improved phase and magnitude reconstruction.

% ============================================================================
\begin{table}[!t]
\centering
\caption{Ablation study on MedleyDB.}
\label{tab:ablation}
{%
\setlength{\tabcolsep}{3.2pt}
\renewcommand{\arraystretch}{1.08}
\begin{tabular}{ccccc cc}
\toprule
MP & ASC & SDR $\uparrow$ & SI-SNR $\uparrow$ & ViSQOL $\uparrow$
& PD $\downarrow$ & LSD $\downarrow$ \\
\midrule
$\checkmark$ & $\checkmark$
& \textbf{17.65} & \textbf{17.27} & \textbf{3.67}
& 13.80 & 0.92 \\

$\checkmark$ & $\times$
& 17.24 & 17.01 & 3.61
& \textbf{13.35} & \textbf{0.87} \\

$\times$ & $\checkmark$
& 17.22 & 16.88 & 3.66
& 13.84 & 0.92 \\
\bottomrule
\end{tabular}
}
\end{table}

\vspace{-3mm}
\subsection{Ablation Study}
\vspace{-1mm}
Table~\ref{tab:ablation} examines the two main design choices of MPASCNet: adaptive spectral compression (ASC) and explicit magnitude-phase modeling (MP). For the ASC ablation, the adaptive spectral compression and reconstruction are replaced with fixed band splitting while retaining the magnitude-phase representation. This variant yields lower PD and LSD, but degrades SDR, SI-SNR, and ViSQOL, suggesting that ASC benefits restoration metrics more than spectral errors.

For the MP ablation, the magnitude-phase representation is replaced with real and imaginary components while retaining adaptive spectral compression. This mainly degrades SDR and SI-SNR, while ViSQOL, PD, and LSD change only slightly. Overall, despite mixed trends in PD and LSD, the complete model achieves the best SDR, SI-SNR, and ViSQOL, indicating stronger overall restoration performance.

% ============================================================================
\begin{figure}[!t]
    \centering
    \includegraphics[width=\columnwidth]{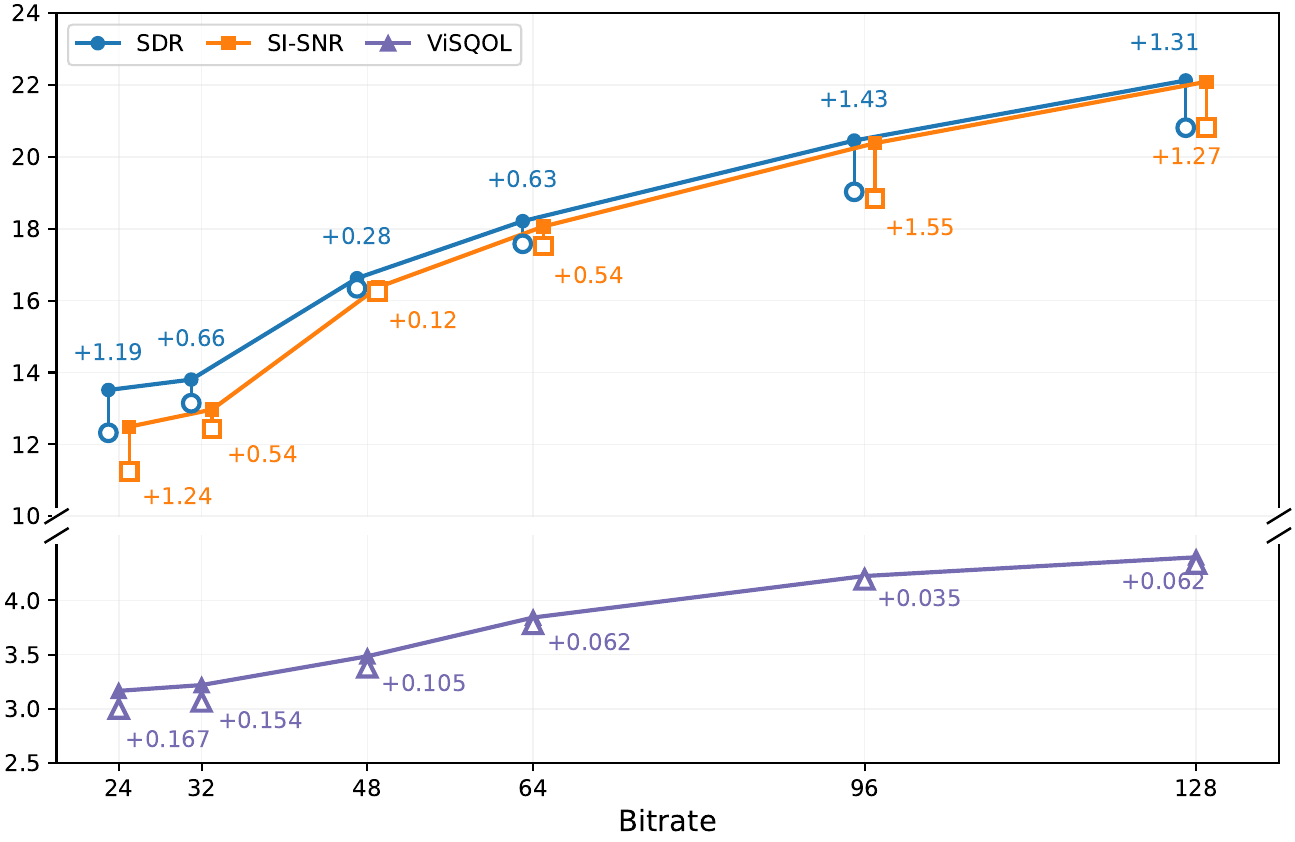}
    \vspace{-3mm}
    \caption{Performance gains of MPASCNet over Apollo under different MP3 bitrates.}
    \label{fig:bitrate}
\end{figure}
\vspace{-3mm}

\subsection{Analysis under Different Bitrates}
\vspace{-1mm}
\label{ssec:bitrate}

Fig.~\ref{fig:bitrate} reports the performance gains of MPASCNet over Apollo across MP3 bitrates. The consistent gains across all evaluated conditions suggest that the proposed representation remains effective across MP3 compression levels.

% ============================================================================
\vspace{-3mm}
\section{Conclusion}
\vspace{-2mm}

We proposed MPASCNet for compressed music restoration, integrating explicit magnitude-phase modeling with input-adaptive spectral feature compression and spectro-temporal modeling on compact spectral tokens. Experiments on MedleyDB show that MPASCNet consistently outperforms the baselines across multiple restoration metrics while using substantially fewer parameters, demonstrating the effectiveness and parameter efficiency of magnitude-phase-aware adaptive spectral compression for compressed music restoration.

\vspace{-3mm}
\section{Acknowledgment}
\vspace{-2mm}
OpenAI ChatGPT was used solely for language refinement. All scientific content, analyses, results, and conclusions were produced and verified by the authors.

\clearpage

\section{Compliance with Ethical Standards}
This study did not require ethical approval under our institutional guidelines.

\bibliographystyle{IEEEbib}
\bibliography{refs}

\end{document}